\documentclass{iau} 
\usepackage{graphicx}

\title[Tests of gravitation with Solar System Objects] 
{Local tests of gravitation with Gaia observations of Solar System Objects}

\author[A. Hees, C. Le Poncin-Lafitte, D. Hestroffer \& P. David]   
{Aur\'elien Hees$^1$, Christophe Le Poncin-Lafitte$^2$, Daniel Hestroffer$^3$ \and Pedro David$^3$
 }

\affiliation{$^1$Department of Physics and Astronomy, University of California, \\ 
Los Angeles, CA 90095, USA \\ email: {\tt ahees@astro.ucla.edu} \\[\affilskip]
$^2$ SYRTE, Observatoire de Paris, PSL Research University, CNRS, Sorbonne Universit\'es, \\
 UPMC Univ. Paris 06, LNE, 61 avenue de l'Observatoire, 75014 Paris, France \\
$^3$ IMCCE, Observatoire de Paris, PSL Research University, CNRS, Sorbonne Universit\'es,\\
 UPMC Univ. Paris 06, Univ. Lille, 77 av. Denfert-Rochereau, 75014 Paris, France}

\pubyear{2017}
\volume{330}  
\jname{Astrometry and Astrophysics in the Gaia sky}
\editors{A. Recio-Blanco, P. de Laverny, A. Brown \& T. Prusti, eds.}
\begin{document}

\maketitle

\begin{abstract}
In this proceeding, we show how observations of Solar System Objects with Gaia can be used to test General Relativity and to constrain modified gravitational theories.  The high number of Solar System objects observed and the variety of their orbital parameters associated with the impressive astrometric accuracy will allow us to perform local tests of General Relativity. In this communication, we present a preliminary sensitivity study of the Gaia observations on dynamical parameters such as the Sun quadrupolar moment and on various extensions to general relativity such as the parametrized post-Newtonian parameters, the fifth force formalism and a violation of Lorentz symmetry parametrized by the Standard-Model extension framework. We take into account the time sequences and the geometry of the observations that are particular to Gaia for its nominal mission (5 years) and for an extended mission (10 years).

\end{abstract}

\firstsection 
\section{Introduction}
Although General Relativity (GR) is currently very well tested (see e.g. \cite[Will 2014]{will14}), there exist strong motivations to pursue searches for modified gravitational theory like e.g. the development of a quantum theory of gravitation, the development of models of dark matter and dark energy, etc. Launched in December 2013, the ESA Gaia mission is scanning regularly the whole celestial sphere once every 6 months providing high precision astrometric data for a huge number ($\approx$ 1 billion) of celestial bodies. In addition to stars, it is also observing Solar System objects (SSOs), in particular asteroids. One can estimate that about 350,000 asteroids will be regularly observed. The high precision astrometry (at sub-mas level) will allow us to perform competitive tests of gravitation and to provide new constraints on alternative theories of gravitation. These constraints will be complementary to the ones existing currently since relying on different bodies, on different type of observations and therefore sensitive to other systematics. In this communication, we report preliminary results of a sensitivity study of Gaia SSOs observations to several modifications of the gravitational theory.

\section{Methodology}
\begin{figure}[htb]
\begin{center}
 \includegraphics[width=.38\textwidth]{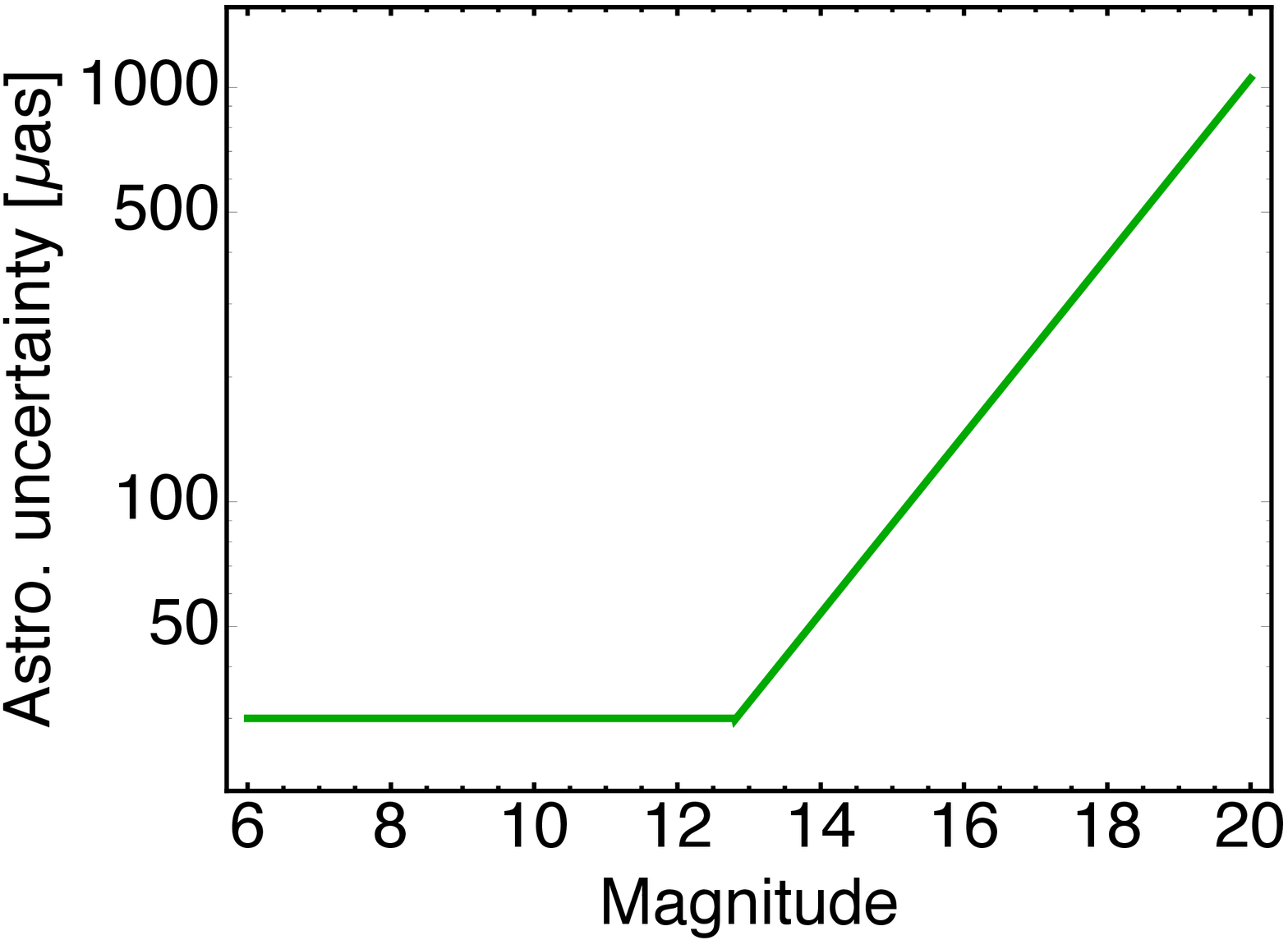} \hspace{1cm}
 \includegraphics[width=.42\textwidth]{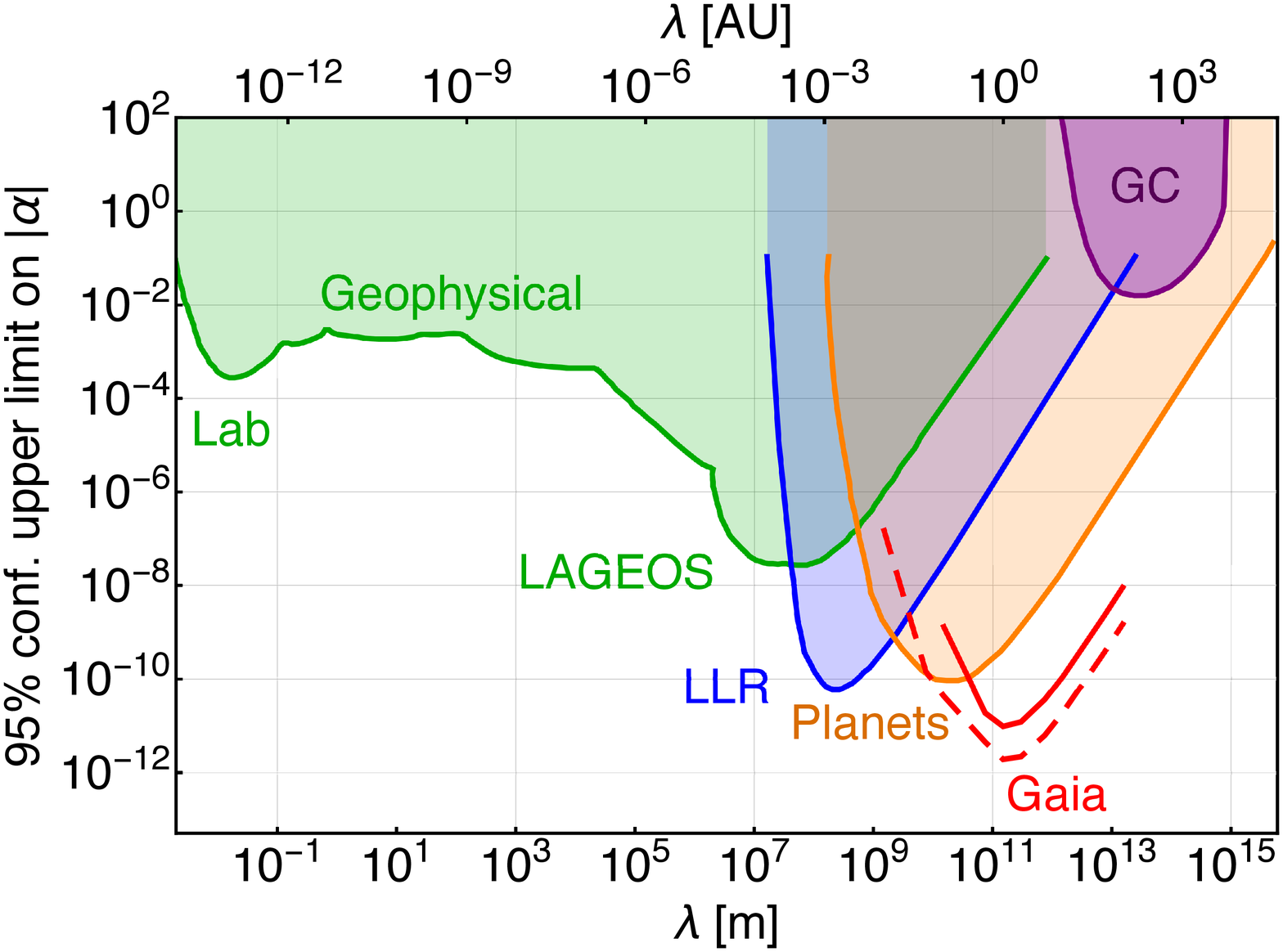} 
 \caption{Left: astrometric uncertainty of SSOs observations as a function of the magnitude. Right: Existing constraints on the fifth force parameters. The  solid Gaia curve corresponds to constraint reachable using 5 years of SSOs observations with Gaia while the dashed Gaia line corresponds to an extended mission.}
   \label{fig:sso}
\end{center}
\end{figure}

In this work, we have considered SSOs from the ASTORB database. A match between their expected trajectories and the Gaia scanning law is performed to find the observation times for each SSO. Two scenarios are considered: (i) the 5 years nominal mission and (ii) a case where the nominal mission is extended by an additional 5 years.  For the 5 years nominal mission, 342,449 SSOs are observed for a total of 20,450,775 observations while for a 10 years mission, this corresponds to 391,518 SSOs for a total of 41,004,470 observations. The astrometric observational uncertainty of each observation depends on the magnitude following a relation which is illustrated on the left panel of Fig.~\ref{fig:sso}.

For each SSO, we integrate the standard post-Newtonian equations of motion in a heliocentric frame. The Sun oblateness $J_2$ is considered and the perturbations from all the planets and the Moon are modeled using the INPOP10e ephemerides (\cite[Fienga et al. 2013]{fienga:2013}). On the other hand, in this preliminary analysis, the mutual interactions between the SSOs are neglected as well as non-gravitational forces. Simultaneously with the equations of motion, we integrate the variational equations to obtain the partial derivatives of the observables with respect to all the estimated parameters, i.e. six initial conditions for each SSO and  global parameters like the Sun $J_2$ and the parameters characterizing the gravitational theory (for a detailed presentation of the method, see~\cite[Hestroffer et al. (2010),  Mouret (2011) and Hees et al. (2015)]{hestro10,mouret11,hees15}). Our sensitivity analysis is based on the Fisher information matrix (or covariance matrix) which gives an estimate of the uncertainty for each parameter as well as correlation coefficients. Therefore, the uncertainties presented in this communication correspond only to statistical uncertainties and our analysis does not include any hypothetical systematics. 

\section{Sensitivity study to various modifications of General Relativity}
In the following, we consider modifications of the gravitational theory that do not impact significantly the light propagation and we focus only on the impact on the orbital dynamics of the SSOs. We report a sensitivity study performed by a global inversion that includes the 6 initial conditions for each of the SSO, the Sun $J_2$ and the parameters characterizing the deviations from GR.

\subsection{The parametrized post-Newtonian framework and the Nordtvedt effect}
The parametrized post-Newtonian (PPN) formalism is a phenomenological framework in which the space-time metric is parametrized by 10 dimensionless coefficients (see \cite[Will 2014]{will14} and references therein). The two most important PPN parameters are $\gamma$ which describes the spatial space-time curvature and $\beta$ which parametrizes the non-linearity in the time component of the space-time metric.  In this analysis, we use $\gamma=1$ since this parameter is better determined by other types of observations like the Shapiro time delay (\cite[Bertotti et al. 2003]{bertotti:2003uq}) and by Gaia itself that will be able to constrain it at the level of $10^{-6}$ by observing light deflection (see e.g. \cite[Mignard \& Klioner 2010]{mignard:2010kx}). In addition, we also consider a violation of the Strong Equivalence Principle which appears in many (if not all) alternative gravitational theories. One effect produced by a violation of the SEP is that the trajectories of self-gravitating bodies depend on their gravitational self-energy $\Omega$. It is characterized by a difference between the gravitational and the inertial mass usually parametrized by the Nordtvedt parameter $\eta$ defined by $m_g=m_i+\eta\frac{\Omega}{c^2}$, where $m_g$ is the gravitational mass and $m_i$ is the inertial mass.

The left part of Table~\ref{tab:PPN} shows the statistical uncertainties that can be reached using SSOs observations from Gaia assuming $J_2$, $\beta$ and $\eta$ to be independent. The uncertainties reachable in the case of the nominal mission are not as competitive as the ones obtained using planetary ephemerides (see e.g. \cite[Fienga et al. 2015]{fienga15}). A 5 years extension of Gaia would result in an improvement of the estimate of $J_2$ by a factor 3, of $\beta$ by a factor 5 and of $\eta$ by one order of magnitude. 

In addition, in metric gravitational theories, the Nordtvedt parameter is univocally related to the PPN parameters through $\eta=4\beta-\gamma-3$ (\cite[Will 2014]{will14}). Assuming this relation improves significantly the estimations of the $\beta$ PPN parameter, as can be seen from the right part of Table~\ref{tab:PPN}. An extended mission would give results as accurate as the ones from the planetary ephemerides.

\begin{table}[htb]
  \begin{center}
  \caption{Statistical uncertainties reachable using Gaia observations to determine  $J_2$, $\beta$ and $\eta$ parameters. In the left part of the table, these three parameters are supposed to be independent while on the right part of the table, $\eta$ is supposed to be directly linked to $\beta$.}
  \label{tab:PPN}
 {\scriptsize
  \begin{tabular}{|r|ccc|cc|}\hline 
 &\multicolumn{3}{c|}{ $J_2$, $\beta$ and $\eta$ independent} &  \multicolumn{2}{c|}{ $\eta=4\beta -4$ } \\ 
 & $\sigma_{J_2}$ & $\sigma_\beta$ & $\sigma_\eta$ & $\sigma_{J_2}$ & $\sigma_\beta$ \\\hline
5 years mission & $4.4 \times 10^{-8}$  & $4 \times 10^{-4}$ & $3\times 10^{-4}$ & $4.1\times 10^{-8}$ & $8 \times 10^{-5}$ \\
10 years mission & $1.5 \times 10^{-8}$ & $9 \times 10^{-5}$ & $3\times 10^{-5}$ & $1.3\times 10^{-8}$ & $8 \times 10^{-6}$ \\\hline
  \end{tabular}
  }
 \end{center}
\end{table}

\subsection{The fifth force formalism}
A fifth force is predicted by several theoretical scenarios motivated by the development of unification theories and models of dark matter. This framework considers deviations from Newtonian gravity in which the gravitational potential takes the form of a Yukawa potential characterized by two parameters: a length $\lambda$ and a strength of interaction  $\alpha$. A summary of the constraints on these two parameters can be found for example in Fig.~2 from \cite[Hees et al (2017)]{hees17}.  The right panel of Fig.~\ref{fig:sso} shows the current constraints on the fifth force parameters as well as the statistical uncertainties expected from Gaia for the nominal and extended missions. The expected results improve the current constraints above 1 astronomical unit. This result is only preliminary and the correlation with the Sun $GM$ still needs to be assessed.

\subsection{A breaking of Lorentz symmetry}
A breaking of Lorentz symmetry is predicted in various unification theories, in a quantum theory of gravity, in non-commutative geometry, \dots The Standard-Model Extension (SME) framework has been developed in order to systematically search for a breaking of Lorentz symmetry in all sectors of physics. In the gravitational sector, at the lowest order, a breaking of Lorentz symmetry is parametrized by a symmetric traceless tensor $\bar s^{\mu\nu}$ (see \cite[Bailey \& Kosteleck\'y 2006]{bailey06}). These coefficients have already been constrained by various observations (for a review, see \cite[Hees et al. 2016]{hees16} and references therein). Nevertheless, current observations did not manage to decorrelate satisfactorily all the SME coefficients. One strong advantage of SSOs observations with Gaia comes from the high number of objects observed and from the variety of their orbital parameters, which allows to decorrelate the SME coefficients. As a result, the estimations of the SME coefficients expected by Gaia, presented in Table~\ref{tab:SME}, are improving the current best constraints by more than one order of magnitude. These results are highly promising. 

\begin{table}[htb]
  \begin{center}
  \caption{Statistical uncertainties reachable using Gaia observations to determine  the SME $\bar s^{\mu\nu}$ coefficients considering a 5 years nominal mission and an extended mission of 10 years.}
  \label{tab:SME}
 {\scriptsize
  \begin{tabular}{|r|cccccccc|}\hline 
 & $\bar s^{XX}-\bar s^{YY}$ & $\bar s^{XX}+\bar s^{YY}-2\bar s^{ZZ}$ & $\bar s^{XY}$ & $\bar s^{XZ}$ & $\bar s^{YZ}$ & $\bar s^{TX}$& $\bar s^{TY}$ & $\bar s^{TZ}$\\
 & [$10^{-12}$] & [$10^{-12}$] & [$10^{-12}$] & [$10^{-12}$] & [$10^{-12}$] & [$10^{-9}$] & [$10^{-9}$] & [$10^{-9}$]\\\hline
5 years mission  & $3.8$  & $6.5$ & $1.7$ & $0.93$ & $1.7$ & $5.7$ & $8.9$ & $16.7$ \\
10  years mission  & $1.5$ & $2.1$ & $0.71$ &$0.38$ & $0.59$ & $1.1$ & $2.1$ & $4.1$ \\\hline
  \end{tabular}
  }
 \end{center}
\end{table}

\section{Conclusion}
In conclusion, Gaia offers a great opportunity to probe fundamental physics by measuring the deflection of light by the gravitational potential of the Sun (\cite[Mignard \& Klioner 2010]{mignard:2010kx}) and by using SSOs observations. The main advantage of Gaia SSOs observations to test GR comes from the wide variety of orbital parameters that can help to decorrelate different parameters. The sensitivity analysis presented in this proceeding relies only on a statistical analysis (no systematic effect has been considered so far). We show that in the PPN framework, an extended mission is expected to provide results competitive with the current best constraints on the PPN parameters while an improvement is expected in the fifth force framework. The more spectacular result is expected within the SME framework where at least an order of magnitude improvement is expected with respect to the current best constraints.

\end{document}